\documentclass[11pt,a4paper]{article}

\usepackage[utf8]{inputenc}
\usepackage[T1]{fontenc}
\usepackage[margin=2.5cm]{geometry}
\usepackage{graphicx}
\usepackage{hyperref}
\usepackage{booktabs}
\usepackage{tabularx}
\usepackage{xcolor}
\usepackage{amsmath}
\usepackage{natbib}
\usepackage{url}
\usepackage{caption}
\usepackage{float}

\hypersetup{colorlinks=true,linkcolor=blue!60!black,citecolor=blue!60!black,urlcolor=blue!60!black}

\title{\textbf{Beyond Human-Readable: Rethinking Software Engineering\\Conventions for the Agentic Development Era}}

\author{
Dmytro Ustynov\\
\textit{Military Institute of Telecommunications}\\
\textit{and Information Technologies (MITIT)}\\
ORCID: 0009-0004-3993-1096
}

\date{April 2026}

\begin{document}

\maketitle

\begin{abstract}
For six decades, software engineering principles have been optimized for a single consumer: the human developer. The rise of agentic AI development, where LLM-based agents autonomously read, write, navigate, and debug codebases, introduces a new primary consumer with fundamentally different constraints. This paper presents a systematic analysis of human-centric conventions under agentic pressure and proposes a key design principle: \textbf{semantic density optimization}---eliminating tokens that carry zero information while preserving tokens that carry high semantic value. We validate this principle through a controlled experiment on log format token economy across four conditions (human-readable, structured, compressed, and tool-assisted compressed), demonstrating a counterintuitive finding: aggressive compression increased total session cost by 67\% despite reducing input tokens by 17\%, because it shifted interpretive burden to the model's reasoning phase. We extend this principle to propose the rehabilitation of classical anti-patterns, introduce the \emph{program skeleton} concept for agentic code navigation, and argue for a fundamental decoupling of semantic intent from human-readable representation.
\end{abstract}


\section{Introduction}

Programming languages have always been translation layers between human intent and machine execution. In every generation, source code was written by humans, for humans, with the compiler serving as the bridge to execution. The entire apparatus of software engineering---naming conventions, design patterns, project structures, SOLID principles, logging formats, commit guidelines---was optimized around human cognition: limited working memory of approximately 7 items~\cite{miller1956}, sequential reading speed, and comprehension through familiar abstractions.

The emergence of agentic AI development tools---Claude Code~\cite{claudecode}, GitHub Copilot~\cite{copilot}, Cursor~\cite{cursor}, Codex CLI~\cite{codex}---disrupts this division. These tools autonomously navigate, read, write, test, and commit code. Reports indicate that some teams now generate 100\% of committed code through AI agents, with humans serving as reviewers~\cite{codescene}. Ronacher describes a workflow where tasks are assigned to an agent and the developer waits for completion~\cite{ronacher2025}.

When the primary consumer shifts from human to machine, the optimization function changes. The relevant constraints become: \textbf{token budgets} (200K--1M tokens, with performance degradation beyond approximately 40\% utilization~\cite{agentstructure}), \textbf{tool call costs} (each file read consumes tokens and context space), \textbf{statistical comprehension} (LLMs understand code through pattern recognition), and \textbf{session impermanence} (each session starts with zero codebase knowledge~\cite{matsen2025}).

An intuitive response to these constraints is compression: make everything shorter. Our experimental findings challenge this intuition and reveal a more nuanced principle. The optimization target is not token minimization but \textbf{semantic density}---the ratio of meaningful information to total tokens. Tokens carrying high semantic value (descriptive names, type annotations, natural language descriptions) are investments that reduce downstream reasoning costs. Tokens carrying zero value (structural boilerplate, ceremonial syntax) are pure waste. Confusing these categories---compressing meaningful content instead of eliminating meaningless overhead---produces counterproductive results, as we demonstrate in Section~\ref{sec:experiment}.

\section{Background and Related Work}

\subsection{Agentic Development Practices}

The practitioner community has accumulated significant experience. Ronacher~\cite{ronacher2025} advocates for Go's simplicity over Python's implicit behaviors, emphasizing that simple code outperforms clever abstractions in agentic contexts. Houston~\cite{houston2025} identifies the tension between human-optimal and agent-optimal project structures, observing that agents struggle with deep package hierarchies. Matsen~\cite{matsen2025} emphasizes context window management as the central resource constraint.

Haupt~\cite{haupt2026} proposes ideas for an agent-oriented programming language, observing that LLMs use fewer tokens for common English words than for shorter programmer abbreviations: Rust's \texttt{fn} is less token-efficient than \texttt{function}. Ronacher's subsequent work~\cite{ronacher2026} identifies whitespace-based indentation as problematic for LLMs and notes that understanding code is becoming more important than writing it.

\subsection{Token-Efficient Data Representation}

The TOON format~\cite{toon} demonstrates 30--60\% token savings over JSON for LLM prompts while maintaining comprehension accuracy~\cite{toonbench}. The New Stack reports that poor data serialization consumes 40--70\% of available tokens in production agent systems~\cite{newstack}. These findings establish that format optimization yields measurable improvements, though our experiment (Section~\ref{sec:experiment}) shows the relationship between input compression and total cost is non-linear.

\subsection{LLM-Oriented Programming Languages}

Quasar~\cite{quasar} proposes a language for LLM code actions with automated parallelization, achieving 42--56\% execution time reduction. Pel~\cite{pel} introduces a Lisp-inspired language for agent orchestration. CoRE~\cite{core} uses the LLM itself as interpreter for structured natural language instructions. These focus on agent orchestration; our work addresses how the entire software engineering stack should evolve.

\subsection{Configuration and Context Engineering}

A study of 2,926 repositories~\cite{configstudy} reveals context files as the dominant agentic configuration mechanism. An ETH Zurich study~\cite{augment} found LLM-generated context files reduced success rates by approximately 3\%, while human-curated files provided marginal 4\% gains---both with over 20\% token overhead, suggesting content quality matters more than presence.

\section{The Semantic Density Principle}

We define the \emph{semantic density} of a software artifact as the ratio of tokens carrying task-relevant meaning to total tokens.

\textbf{High-density tokens} include: descriptive function/variable names, type annotations, docstrings, business logic expressions, error messages with diagnostic context.

\textbf{Zero-information tokens} include: structural boilerplate (class declarations, DI wiring), syntactic ceremony (redundant access modifiers, interface declarations for single implementations), and framework-mandated scaffolding.

\textbf{The principle:} Optimizing for agentic consumption requires maximizing semantic density---eliminating zero-information tokens while preserving high-information tokens. Critically, compressing high-information tokens is counterproductive: it reduces input tokens but increases reasoning tokens, as the model must reconstruct lost semantics. This is validated experimentally in Section~\ref{sec:experiment}.

\section{A Taxonomy of Conventions Under Pressure}

\begin{figure}[H]
\centering
\includegraphics[width=\textwidth]{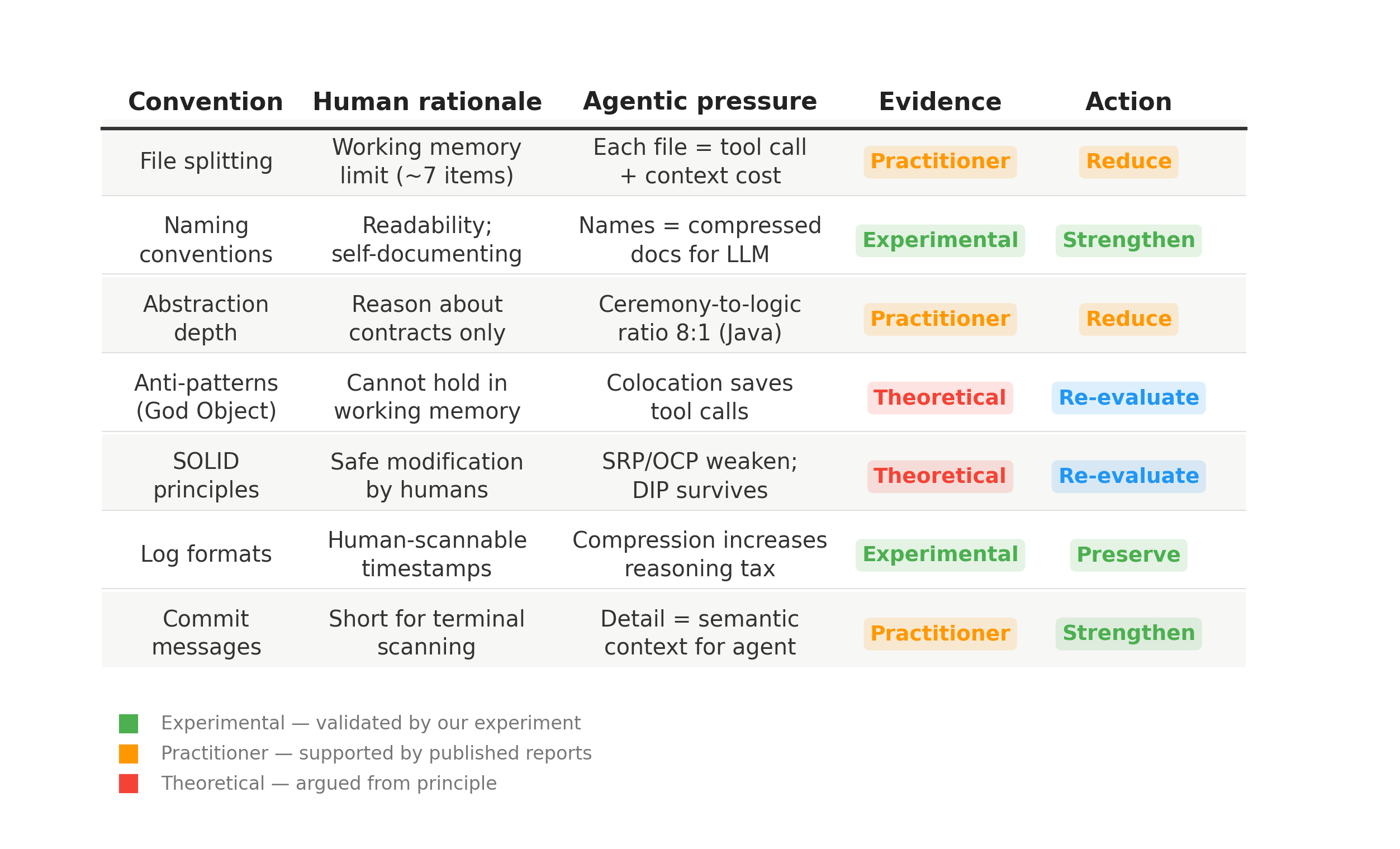}
\caption{Taxonomy of software engineering conventions under agentic pressure. Evidence levels indicate the strength of support for each recommendation.}
\label{fig:taxonomy}
\end{figure}

We identify seven categories of conventions whose cost-benefit profile shifts under agentic development, summarized in Figure~\ref{fig:taxonomy}.

\subsection{File Splitting and Project Structure}

Humans split logic across files because each must fit in working memory. For agents, each file read is a tool call consuming tokens and context. Houston~\cite{houston2025} recommends consolidating micro-packages into broader modules. Vertical slice architecture, where all code for a feature lives together, emerges as naturally agent-friendly.

\textbf{Recommendation:} File separation should be driven by deployment boundaries, testability, and concurrent development---not cognitive limits on file size.

\subsection{Naming Conventions and Semantic Density}

Unlike most conventions, meaningful naming \emph{increases} in value. \texttt{VerifyOrderByAvailableAmount} is compressed documentation the model interprets semantically. Our experiment (Section~\ref{sec:experiment}) demonstrates the equivalent for logs: abbreviated labels saved input tokens but increased reasoning costs.

\textbf{Recommendation:} Strengthen naming conventions. Token economy gains come from eliminating zero-information structural tokens, not high-information semantic tokens.

\subsection{Abstraction Depth and Ceremony}

We introduce the \emph{ceremony-to-logic ratio} as an informal metric. A typical Java Spring Boot endpoint spans 8+ files containing $\sim$170 lines, of which $\sim$18 are business logic---a ratio exceeding 8:1. For agents, this means 9 tool calls and hundreds of zero-information tokens. Ronacher~\cite{ronacher2025}: ``Have the agent do `the dumbest possible thing that will work.' ''

\textbf{Recommendation:} Flat call chains with well-named functions over deep hierarchies. Abstraction depth should be justified by concrete benefits, not cognitive management.

\subsection{Classical Anti-Patterns Under Re-Evaluation}

We do not argue that anti-patterns become best practices. Rather, their cost-benefit profile shifts, and the degree is an open empirical question subject to the caveats in Section~\ref{sec:limitations}.

An 800-line file costs 5,000--10,000 tokens (trivial for 200K context) and one tool call. The same logic in 15 files: 15 tool calls, 20,000+ tokens with boilerplate. However, the ``Lost in the Middle'' phenomenon~\cite{liu2024} demonstrates degraded attention for centrally-positioned information, directly challenging large-file rehabilitation. A God Object offers colocation benefits but faces training distribution concerns (Section~\ref{sec:limitations}).

\subsection{SOLID Principles}

SRP's cognitive argument weakens; testability survives. OCP's cost (complex extension hierarchies) increases relative to its benefit. DIP may \emph{increase} in value, though specific mechanisms (annotation-driven DI containers) need revision. DRY's rationale weakens when agents can reliably find and update all copies, though large-scale repetition remains harmful.

\subsection{Logging and Observability Formats}

See Section~\ref{sec:experiment}. Key finding: human-readable formats produce lower total task costs than compressed formats at moderate scale.

\subsection{Git Commit Messages}

The 50-character convention serves terminal scanning. Agents benefit from rich messages capturing reasoning, alternatives, and constraints. This aligns with semantic density: detailed commits are high-density content saving the agent from re-reading diffs.

\section{The Program Skeleton}
\label{sec:skeleton}

We propose the \textbf{program skeleton} as a new software artifact category for agentic consumption. We recommend the filename \texttt{CODEMAP.md} by convention.

\textbf{Contains:} module topology, entry points, call chains with one-line descriptions, function signatures with docstrings, data flow summary, file locations.

\textbf{Omits:} implementation bodies, imports, structural boilerplate, framework ceremony.

The skeleton is \emph{not} an AST (which is typically larger than source code) but a lossy semantic compression preserving navigation. It differs from LSP (Language Server Protocol) queries: LSP provides dynamic per-query access, while the skeleton provides batch orientation in a single read. The two are complementary: skeleton for the map, LSP for on-demand detail.

IDE vendors are uniquely positioned to produce skeletons: they already perform the required semantic analysis (call graphs, type resolution, usage tracking). The missing piece is serialization for LLM consumption rather than GUI rendering. A committed skeleton would provide persistent structural knowledge across agent sessions, eliminating the current pattern where each session rebuilds understanding from scratch.

\section{Experiment: Log Format Token Economy}
\label{sec:experiment}

\subsection{Methodology}

We generated 200 realistic log events simulating a 30-minute e-commerce application window, covering HTTP requests, database queries, authentication, business logic, errors with stack traces, and warnings. The same events were encoded in four conditions:

\begin{itemize}
\item \textbf{Format A --- Human-Readable:} Natural language, formatted timestamps, full names.
\item \textbf{Format B --- Structured:} Pipe-delimited, Unix timestamps, full names, key-value pairs.
\item \textbf{Format C --- Compressed:} Abbreviated service/event codes, schema header.
\item \textbf{Format D --- C + Decoder Tool:} Format C with a Python script for on-demand translation via targeted queries.
\end{itemize}

File-level tokens were measured with cl100k\_base (tiktoken). Comprehension was tested by presenting five identical diagnostic questions to \texttt{claude-sonnet-4-6} in isolated sessions (one per format, no cross-contamination), each starting fresh at 15.1k baseline tokens. No extended thinking was used.

\subsection{Results}

\subsubsection{File-Level Token Counts}

\begin{table}[H]
\centering
\begin{tabular}{lrrrr}
\toprule
\textbf{Format} & \textbf{Tokens} & \textbf{Lines} & \textbf{Tok/Line} & \textbf{$\Delta$ vs.\ A} \\
\midrule
A --- Human-Readable & 8{,}072 & 247 & 32.68 & --- \\
B --- Structured & 7{,}106 & 200 & 35.53 & $-$12.0\% \\
C --- Compressed & 6{,}695 & 203 & 32.98 & $-$17.1\% \\
\bottomrule
\end{tabular}
\caption{File-level token counts (cl100k\_base tokenizer).}
\label{tab:filetokens}
\end{table}

\subsubsection{Session-Level Results}

\begin{table}[H]
\centering
\begin{tabular}{lrrrr}
\toprule
\textbf{Metric} & \textbf{A (Human)} & \textbf{B (Struct.)} & \textbf{C (Compr.)} & \textbf{D (C+Tool)} \\
\midrule
Session tokens (Messages) & 18.9k & 24.0k & 31.6k & 28.3k \\
Wall-clock time & 1m 36s & 5m 24s & 7m 00s & 4m 05s \\
Tool calls & 1 & 1 & 1 & $\sim$5--7 \\
Correctness (of 5) & 5/5 & 5/5 & 5/5 & 5/5 \\
HIGH confidence & 4/5 & 4/5 & 3/5 & 4/5 \\
\bottomrule
\end{tabular}
\caption{Session-level results across four experimental conditions.}
\label{tab:session}
\end{table}

\subsubsection{The Compression Paradox}

\begin{figure}[H]
\centering
\includegraphics[width=0.95\textwidth]{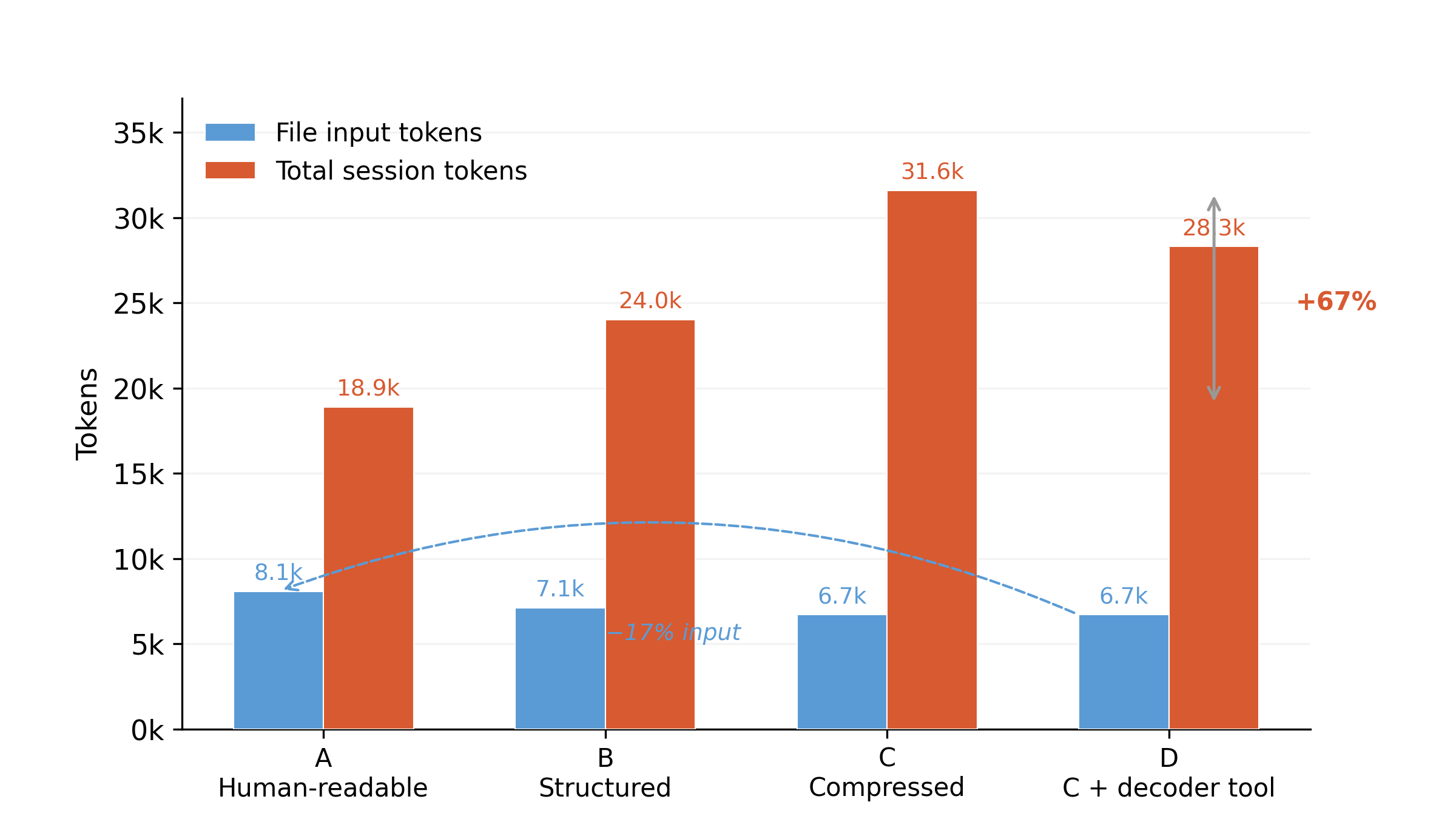}
\caption{The compression paradox: input tokens decrease with compression (blue), but total session tokens increase (orange). Annotated: 17\% input savings produced 67\% session cost increase.}
\label{fig:paradox}
\end{figure}

The most striking finding is the inversion between file-level and session-level costs (Figure~\ref{fig:paradox}). Format~C achieved 17.1\% input reduction but produced 67.2\% session cost increase. The session-to-file ratio grew from 2.3$\times$ (A) to 4.7$\times$ (C), indicating progressively more reasoning tokens as input density increased.

\subsection{Analysis}

\subsubsection{The Reasoning Tax}

\begin{figure}[H]
\centering
\includegraphics[width=0.95\textwidth]{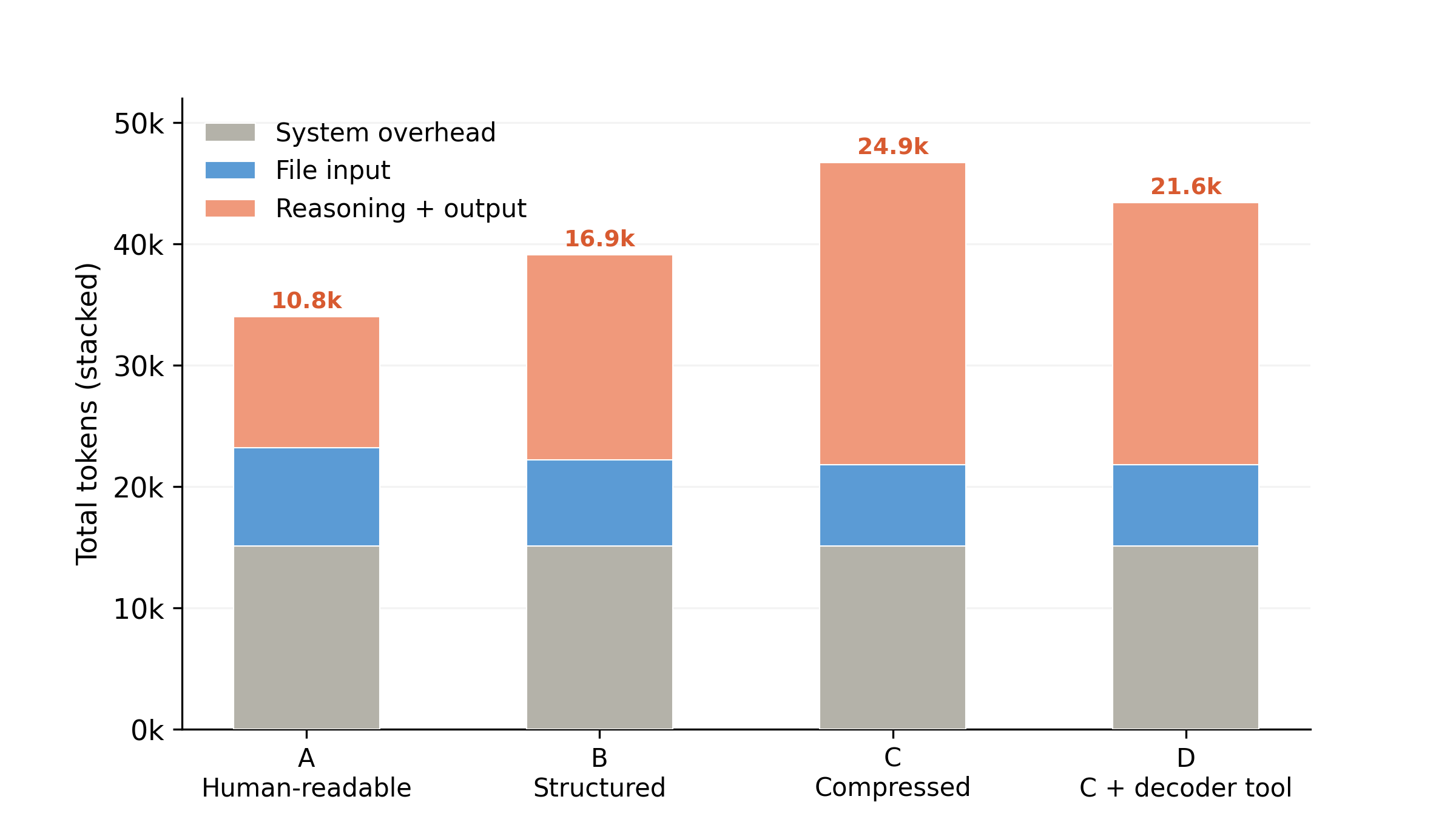}
\caption{Stacked token breakdown. The orange ``reasoning + output'' component grows as input compression increases, demonstrating the reasoning tax. Numbers above bars show reasoning tokens.}
\label{fig:tax}
\end{figure}

Total task cost is $\text{input} + \text{reasoning} + \text{output}$. Compressing meaningful information shifts burden to reasoning (Figure~\ref{fig:tax}).

In Format~A, \texttt{"Payment failed for order \#4521: insufficient funds"} is immediately interpretable. In Format~C, \texttt{|E|PS|pf|o=4521|rs=insuf\_funds} requires the model to decode each abbreviation and explain decodings in output. The model explicitly noted ``undefined service codes'' in Format~C, spending tokens puzzling over abbreviations self-evident in natural language.

\subsubsection{Tool Overhead}

Format~D reduced reasoning tax versus raw compressed (28.3k vs.\ 31.6k) but introduced tool-calling overhead: 5--7 invocations, an execution error (\texttt{python} vs.\ \texttt{python3}), and an alternative-approach detour. At 200 lines, fixed per-call overhead exceeds selective decompression savings. The crossover point likely occurs at volumes exceeding context window capacity.

\subsubsection{Implications and Scope}

The experiment validates semantic density for \textbf{data retrieval tasks}. Extension to code architecture (file consolidation, anti-pattern rehabilitation) remains theoretically motivated but not experimentally validated. Log analysis is linear retrieval; code modification is synthesis and dependency mapping. The experiments in Section~\ref{sec:future} are designed to address this gap.

\section{Language-Level Analysis}

\subsection{Java: High Ceremony-to-Logic Ratio}

A Spring Boot REST endpoint spanning 8+ files with $\sim$170 lines yields $\sim$18 lines of business logic. The 150+ remaining lines are zero-information tokens from the semantic density perspective. Stack traces compound the problem: 50+ lines per exception, 80\% framework internals.

\subsection{Go as Contrast}

Go's simplicity aligns with semantic density: structural interfaces, uniform formatting, explicit error handling, low churn. The same endpoint requires 2--3 files with a ceremony-to-logic ratio near 2:1.

\subsection{Framework-Level Ceremony}

Angular generates four files per component. Ceremony ratios are cumulative: language $\times$ framework $\times$ architecture produces multiplicative overhead.

\section{Discussion}

\subsection{The Decoupling Thesis}

The underlying pattern is a decoupling of semantic intent from human-readable representation. Source maps, decompilers, and IDE indexes are existing instances. The skeleton and dual-format commit message extend this pattern. ``Machine-optimized'' does not mean ``compressed''---it means ``semantically dense,'' which often aligns with human readability.

\subsection{When Compression Becomes Necessary}

At scale exceeding context capacity, compression or selective access becomes necessary. The skeleton addresses this for code; the decoder tool addresses it for logs. The principle remains: preserve semantic density within whatever is loaded.

\subsection{Implications for Education}

Curricula should teach principles behind conventions separately from specific implementations, introduce semantic density as a design consideration, and critically examine which practices are universal versus human-cognitive optimizations.

\subsection{Limitations and Open Challenges}
\label{sec:limitations}

\subsubsection{Training Distribution Paradox}

LLMs are proficient with SOLID patterns because they were trained on millions of examples. Agent-optimal structures (consolidated files, God Objects) are underrepresented in training data. Deviating may increase hallucination and decrease quality---``distributional drift.'' Training corpora include non-enterprise code, and future training will include agent-generated code, but neither eliminates the concern.

\subsubsection{Attention Degradation}

The ``Lost in the Middle'' phenomenon~\cite{liu2024} shows degraded retrieval for centrally-positioned information, directly challenging large-file rehabilitation. The skeleton partially mitigates this via explicit pointers, but the degree of mitigation is untested.

\subsubsection{Human Reviewer Cognitive Debt}

Optimizing for agents transfers cognitive load to human reviewers. A convention is optimal only if total system cost (agent development + human review) is minimized. The projection layer concept addresses this, but such tools are nascent. Consolidation should be calibrated to the team's review workflow.

\subsubsection{Proxy Gap}

Our experiment validates semantic density for data retrieval. Architectural claims are supported by practitioner observations~\cite{ronacher2025,matsen2025,houston2025} but not controlled experiment. Until the experiments in Section~\ref{sec:future} are conducted, architectural recommendations should be understood as theoretically motivated hypotheses.

\subsubsection{Methodological}

Single model (\texttt{claude-sonnet-4-6}), single dataset (200 events). The crossover point for compression is not established. Agent capabilities evolve rapidly.

\section{Future Work}
\label{sec:future}

\textbf{Scale-dependent optimization:} Systematic measurement of the crossover point where compressed format with tool-assisted decompression becomes more efficient than human-readable.

\textbf{Skeleton evaluation:} Controlled comparison with and without \texttt{CODEMAP.md}, measuring tool calls, tokens, time, and correctness across three conditions (no skeleton, human-curated, agent-generated).

\textbf{File consolidation:} Identical functionality in multi-file versus consolidated architecture, directly testing whether eliminating file-level ceremony improves agent efficiency.

\textbf{Ceremony-to-logic formalization:} Classify AST nodes as ``ceremony'' (imports, access modifiers, single-implementation interfaces) versus ``logic'' (conditionals, business calls, transformations). Compute across a corpus to establish baselines.

\textbf{LLM-native language design:} Language designed for maximal semantic density: minimal ceremony, rich naming as first-class design, integrated skeleton generation, dual human/machine representation.

\section{Conclusion}

Our central contribution is the semantic density principle, validated through a controlled experiment demonstrating that aggressive compression is counterproductive for LLM task performance. The principle redirects optimization toward elimination of zero-information structural overhead.

We propose the program skeleton for agent-efficient navigation and argue for conditional re-evaluation of anti-patterns predicated on human cognitive limits. These architectural proposals are theoretically motivated and require the empirical validation described above. The training distribution paradox, attention degradation, and reviewer cognitive debt are genuine constraints.

The finding we are most confident in: the optimal representation for both humans and machines shares high semantic density. Well-named functions, clear documentation, and meaningful structure benefit both. The divergence lies in structural organization---humans need small files and visual formatting; machines need consolidated access and navigational indexes. Reconciling these through projection layers and dual-format tooling is the central question for the emerging field of LLM-oriented software engineering.

\bibliographystyle{plainnat}

\end{document}